\documentclass{article}
\usepackage{psfig}
\usepackage{amsmath}
\usepackage{amssymb}

\begin{document}

\markboth{Luca Capriotti}
{A closed-form approximation for likelihood functions  ...}

\title{A CLOSED-FORM APPROXIMATION OF LIKELIHOOD FUNCTIONS FOR 
DISCRETELY SAMPLED DIFFUSIONS: \\ THE EXPONENT EXPANSION}

\author{Luca Capriotti \\
{Global Modelling and Analytics Group, }\\ 
Credit Suisse, Investment Banking Division \\ 
One Cabot Square, London, E14 4QJ, United Kingdom \\
e.mail: \texttt{luca.capriotti@credit-suisse.com} }

\maketitle

\begin{abstract}
In this paper we discuss a closed-form approximation 
of the likelihood functions of an arbitrary diffusion process. 
The approximation is based on an exponential ansatz of the 
transition probability for a finite time step $\Delta t$, 
and a series expansion of the deviation of its logarithm from that of a Gaussian
distribution. Through this procedure, dubbed {\em exponent expansion}, 
the transition probability is obtained as a power series in $\Delta t$.
This becomes asymptotically exact 
if an increasing number of terms is included, and provides remarkably accurate 
results even when truncated to the first few (say 3) terms. The coefficients of such 
expansion can be determined straightforwardly through a recursion, 
and involve simple one-dimensional integrals. 

We present several examples of financial interest, and we compare our
results with the state-of-the-art approximation of discretely
sampled diffusions [A\"it-Sahalia, {\it Journal of Finance} {\bf 54}, 1361 (1999)]. 
We find that the exponent expansion provides a similar accuracy 
in most of the cases, but a better behavior in the low-volatility regime.
Furthermore the implementation of the present approach turns out to be 
simpler.

Within the functional integration framework the exponent expansion allows 
one to obtain remarkably good approximations of the pricing kernels 
of financial derivatives. This is illustrated with the application 
to simple path-dependent interest rate derivatives.
Finally we discuss how these results can also be used  
to increase the efficiency of numerical (both deterministic and stochastic) 
approaches to derivative pricing.

\end{abstract}

\section{INTRODUCTION}

Continuous-time diffusion processes is the basis of much of the  
modeling work performed every day in Finance and Economics, from portfolio
optimization and econometric applications, to contingent claim pricing. 
Indeed, since Bachelier's 1900 doctoral thesis on the 
dynamics of stock prices \cite{bachelier}, 
many economic variables subject to unpredictable fluctuations, 
have been modeled by stochastic differential equations of the form
\begin{equation}
dY_t = \mu_y(Y_t) dt + \sigma_y(Y_t) dW_t~. 
\label{diffusionY}
\end{equation}
Here $\mu_y(y)$ is the drift, describing a deterministic trend,
and $\sigma_y(y) \ge 0$ is the volatility function, describing the level of randomness
introduced by the Wiener process (i.e., white noise), $dW_t$. The main reason for the 
popularity of this class of models is probably that in continuous time one can perform 
analytic calculations using the instruments of stochastic calculus, 
and the powerful framework of partial differential equations. 

In particular, for the few cases for which the process (\ref{diffusionY}) is exactly solvable,
one can derive closed-form solutions for the associated transition probability. The latter 
contains all the statistical properties of the financial quantity modeled
by the diffusion, and can be exploited in a variety of ways, including 
the derivation of no-arbitrage prices for financial derivatives in complete markets. 
The milestone results derived by Black, Scholes and Merton \cite{bs1,bs2}, Cox, Ingersoll and Ross \cite{cir}, 
or Vasicek \cite{vasicek},  are among the most significant examples of the amount of 
progress in Economics that has been 
done using integrable continuous-time diffusion processes.

Nonetheless, an accurate description of the market observables requires in general more
sophisticated models than those for which an analytic solution is available. These 
are usually tackled by means of numerical schemes ultimately relying on a 
discretization of the diffusion, obtained by replacing the infinitesimal time $dt$ with 
a finite time step, $\Delta t$.
These approaches typically involve either solving numerically a
Kolmogorov partial differential equation, or a Monte Carlo sampling of the diffusive paths.
The approximate results obtained in this way 
become exact only approaching the limit $\Delta t \to 0 $, and this can be done 
with some computational effort.  In addition, a drawback of these numerical methods, more specific
to econometric applications, is that they do not  produce closed-form 
expressions for the transition probability. These are crucial for 
maximum-likelihood estimations of the parameters, say $\theta$, of model diffusions.
In fact, economic data are generally available on discrete sets of
observations, usually well spaced in time, say weekly or
monthly. As a result, only if the transition probability, $\rho(X_{(l+1)\Delta t},\Delta t| X_{l\Delta t}; \theta)$,
associated with each time interval $\Delta t$ of the series $t = l \Delta t$  ($l =1,\ldots n$),
is known in closed form, the maximum likelihood function, 
\begin{equation}
l_n(\theta) = \frac{1}{n} \sum_{l=1}^n \log{\rho(X_{(l+1) \Delta t}, l \Delta t| X_{l \Delta t} ; \theta)}~,
\label{mlf}
\end{equation}
can be analytically maximized over $\theta$. If this is not the case, one has to repeat
the numerical calculation of the transition probability for every value of $\theta$
needed for the determination of the maximum, e.g. by means of an optimization algorithm.
This can be clearly very time consuming.

Motivated by this difficulty, A\"it-Sahalia recently proposed a method to approximate the
transition probability for one dimensional diffusions \cite{aitsahalia} by means of a Hermite
polynomial expansion. This was applied to a variety of test cases, and the accuracy of the method
was clearly demonstrated. 

In this paper, we utilize the {\em exponent expansion} -- a technique introduced in chemical physics 
by Makri and Miller \cite{makri} -- to derive a closed-form short-time approximation of the transition 
probability of the diffusion process (\ref{diffusionY}).
The aim is to obtain an analytic approximation which is
as accurate as possible for a time step $\Delta t $ as large as possible.  
On one hand, this allows one to derive approximations of financial quantities
that are very accurate even for sizable values of the time step, and to derive
closed-form expression for the maximum likelihood function (\ref{mlf}). On the other, it allows a 
reduction of the computational burden of numerical schemes 
as the limit $\Delta t \to 0 $ can be achieved with larger time steps, i.e., with
less calculations. Our approach is similar in spirit to the one of Ref.~\cite{aitsahalia}, reviewed
in Section \ref{aitreview},
but it overcomes some of its shortcomings, and it is of simpler implementation.
In particular, the coefficients of the exponent expansion can be expressed in terms
of one dimensional integrals that can be easily calculated numerically.
In addition,  we show how we can apply our approach to any sufficiently regular
volatility function, even when the latter does not have an analytic expression but it is
specified numerically through an interpolation procedure, 
as it is the typically the case for  local volatility models.

The possibility to use Makri and Miller's technique to derive approximations 
of the transition probability was originally hinted by Bennati 
{\em et al.} in Ref.~\cite{rosaclot1,rosaclot2}. Here we explore this route,
giving derivations for a generic diffusion process with state-dependent 
drift and volatility, and we study the reliability of the exponent expansion 
by applying it to several test problems of financial interest. 

Through the exponent expansion, the transition probability is obtained as a power 
series in $\Delta t$ which becomes asymptotically exact if an increasing number 
of terms is included, and provides remarkably accurate results even when truncated
to the first few (say, $n=3$) terms. Two derivations are offered, the first by means of Kolmogorov's 
forward equation \cite{shreve} (Sec.~\ref{expa}), 
and the second introducing a slightly different formalism (Sec.~\ref{alternative}).
The latter, once the problem is formulated in terms of 
Feynman's path integrals \cite{pireference1,pireference2}, allows the generalization of the exponent expansion to
the calculation of the pricing kernel of financial derivatives whose underlying follows the
considered diffusion. This allows in turn the derivation of simple approximations for 
the price of such contingent claims (Sec.~\ref{prickern}). 
In Sections \ref{app1} and \ref{app2}, we illustrate the exponent expansion through the application to the Vasicek, the Cox-Ingersoll-Ross, and the Constant Elasticity of Variance models, and in Section \ref{montecarlo} 
we discuss its application to Monte Carlo and deterministic numerical methods 
within the path integral framework \cite{montagna1,montagna2,baaquiep,matacz,dash,rosaclot2}. 
Finally, we draw our conclusions, and we discuss future developments in Section \ref{conclusion}.

\section{TRANSITION PROBABILITIES OF DIFFUSION PROCESSES}

Let us consider the problem of estimating the transition probability,
$\rho_y(y,\Delta t|y_0)$,
associated with one-dimensional continuous-time diffusion processes of the form (\ref{diffusionY}).
This represents the likelihood
that the random walker following the process (\ref{diffusionY}) ends up in 
the position $y$ at time $t=\Delta t$, given that it was in $y_0$ at time $t = 0$, and
satisfies the Kolmogorov forward (or Fokker-Planck) equation \cite{shreve}:
\begin{equation}
\partial_t \rho_y(y,\Delta t|y_0)  = \left[ -\partial_y \mu_y(x) +
\frac{1}{2} \partial^2_y \sigma_y(y)^2 \right] \rho_y(y,\Delta t|y_0)~.
\label{kolmogorovgen}
\end{equation}
In this Section we will consider two short-time approximations of the transition
probability above that can be derived in closed form. We will start
by reviewing the Hermite polynomial expansion, recently introduced by A\"it-Sahalia ~\cite{aitsahalia}, 
and then describe the exponent expansion.

\subsection{Review of the Hermite polynomials expansion}
\label{aitreview}

The first step of A\"it-Sahalia's derivation is  to transform the 
original process in an auxiliary one, say $X_t$, with constant volatility 
$\sigma_x$. Following Ref.~\cite{aitsahalia}, this
can be achieved in general through the following integral transformations
\begin{equation}
X_t = \gamma(Y_t) \equiv \pm \sigma_x \int^{Y_t}  \frac{dz}{\sigma_y(z)}~,
\label{inttransf}
\end{equation}
where the choice of the sign is just a matter of
convenience depending on the specific problem considered.
The latter relation defines a one to one mapping between the $X_t$ and $Y_t$
processes as the condition $\sigma_y(z) \ge 0$ ensures that the function
$x = \gamma(y)$ defined by (\ref{inttransf}) is monotonic, and therefore
invertible. \footnote{For a discussion of the regularity conditions on the drift and volatility
functions see e.g., Ref.~\cite{aitsahalia}.} A straightforward application of Ito's Lemma \cite{shreve}
allows one to write the diffusion process followed by $X_t$ as
\begin{equation}
dX_t = \mu_x(X_t) dt + \sigma_x dW_t~,
\label{processX}
\end{equation}
with
\begin{equation}
        \mu_x(x) = \pm \sigma_x \left[ \, \frac{\mu_y(\gamma^{-1}(x))}{\sigma_y(\gamma^{-1}(x))} - \frac{1}{2}\frac{\partial \sigma_y}{\partial y}(\gamma^{-1}(x))\right]~,
\label{drift}
\end{equation}
where $y = \gamma^{-1}(x)$ is the inverse of the transformation (\ref{inttransf}).

Using Hermite polynomials, it is possible to show \cite{aitsahalia}, that a short time
approximation of the transition probability can be expressed up to order $N$ as
\begin{equation}\label{aitsa1}
\rho_x(x,\Delta t|x_0) = \Delta t^{-1/2} \phi(\frac{x-x_0}{\Delta t^{1/2}}) 
\exp{\big(\int_{x_0}^x dz \mu_x(z)\big)} \sum_{n=0}^N c_n(x,x_0) \frac{(\Delta t)^n}{n!}~,
\end{equation}
where $\phi(x)$ is a standard normal distribution.
Here the coefficients $c_n(x,x_0)$ can be derived by solving the recursive equation
\begin{equation}\label{aitsa2}
c_n(x,x_0) = n (x-x_0)^{-n} \int_{x_0}^x dz (z-x_0)^{n-1} 
\left(\lambda(z) c_{n-1} (z,x_0)+\frac{1}{2}\partial_z^2 c_{n-1}(z,x_0)\right)~,
\end{equation}
for $n>0$, with $c_0(x,x_0) \equiv 1$, and
\begin{equation}\label{aitsa4}
\lambda(x) = -\frac{1}{2} \left(\mu_x(x)^2 +\partial_x \mu_x(x) \right).
\end{equation}

Finally, the transition probability for the process $Y_t$ can be determined through the Jacobian of the transformation
(\ref{inttransf}) giving
\begin{equation}
\rho_y(y,\Delta t|y_0) = \sigma_x \frac{\rho_x(\gamma(y),\Delta t| x_0)}{\sigma_y (y)}~.
\label{aitsa5}
\end{equation}

The expansion above  turns out to provide very accurate results in a variety of test cases
for which the exact expression of the transition probability is available. Some of these
examples will be considered in detail in the following when we will compare the results
obtained with the approximation Eqs.~(\ref{aitsa1})-(\ref{aitsa5}), with those
given by the exponent expansion described in this paper, and introduced in the 
following Section.

\subsection{The Exponent Expansion}
\label{expa}

In this Section, we derive the exponent expansion for the 
transition probability
for the process (\ref{diffusionY}), $\rho_y(y,\Delta t|y_0)$.
In order to make the derivation easier, it is convenient to 
transform the original process in the constant volatility one 
Eq.~(\ref{processX}) by means of the transformation (\ref{inttransf}).

For $\Delta t\to 0$ the transition probability for $X_t$ is dominated
by the diffusive Gaussian component and therefore reads:
\begin{equation}
\rho_x(x,\Delta T |x_0) = \left( \frac{1}{2\pi \sigma_x^2 \Delta t}\right)^{1/2} 
\exp{\left(-\frac{(x-x_0)^2}{2\sigma_x^2\Delta t}\right)}~.
\end{equation}
Guided by this observation, in order to find an expression for the transition probability associated with Eq.~(\ref{processX}) 
which is accurate for a time $\Delta t$ as long as possible, we make the following {\em ansatz}:
\begin{equation}
\rho_x(x,\Delta t|x_0) = \frac{1}{\sqrt{2\pi\sigma_x^2\Delta t}} 
\exp {\left[ -\frac{(x-x_0)^2}{2\sigma_x^2 \Delta t}-W(x,x_0,\Delta t)\right]}~.
\label{ansatz}
\end{equation}

Such transition probability must satisfy the Kolmogorov forward equation \cite{shreve}:
\begin{equation}
\partial_t \rho(x,\Delta t|x_0)  = \left[ -\partial_x \mu_x(x) + 
\frac{1}{2}\sigma_x^2 \partial^2_x \right] \rho_x(x,\Delta t|x_0)~.
\label{kolmogorov}
\end{equation}
Note that since the auxiliary process $X_t$ has a constant volatility, the latter 
equation is mathematically much simpler than the one 
for the process $Y_t$ (\ref{kolmogorovgen}).
Equation (\ref{kolmogorov}) implies in turn that the function $W(x,x_0,t)$ satisfies the relation:
\begin{equation}
\partial_t W = - \mu_x \partial_x W
+ \frac {1}{2} \sigma_x^2 \partial^2_x W 
- \frac {1}{2} \sigma_x^2 \left( \partial_x W \right)^2
+ \partial_x \mu_x 
- \frac{x-x_0}{\Delta t} \,\left( \partial_x W+\frac{\mu_x}{\sigma_x^2} \right)~.
\label{eqW}
\end{equation}
Expanding $W(x,x_0,t)$ in powers of $\Delta t$,
\begin{equation}
W(x,x_0,\Delta t) = \sum_{n=0}^{\infty} W_n(x,x_0)\,\Delta t^n~,
\label{w}
\end{equation}
substituting it in Eq.~(\ref{eqW}), and equating
equal powers of $\Delta t$ leads in a straightforward way to a
decoupled equation for the order zero in $\Delta t$ giving
\begin{equation}
W_0(x,x_0) = - \frac{1}{\sigma_x^2}\int_{x_0}^{x} dz \,\mu_x(z)~,
\label{w0}
\end{equation}
and to the following set of recursive differential equations:
\begin{eqnarray}
(n+1)W_{n+1} &=& - (x-x_0)\partial_x W_{n+1} + 
\left[ \frac{1}{2}\,\sigma_x^2 \partial^2_x -\mu_x\partial_x \right] W_n \nonumber \\
&-& \frac{1}{2}\sigma_x^2 \sum_{m = 0}^{m = n}\partial_x W_{m}\partial_x W_{n-m} 
+ \delta_{n,0} \partial_x \mu_x~.
\label{makrieq}
\end{eqnarray}
In particular, for $n=0,1,2$ Eqs.~(\ref{makrieq}) read:
\begin{eqnarray}
	W_1(x,x_0) &=& -(x-x_0)\,\partial_x W_1(x,x_0)+ \left[ \frac{1}{2\sigma_x^2} \mu_x(x)^2 + \frac{1}{2}\partial_x \mu_x(x) \right]~,\\
  2W_2(x,x_0) &=& -(x-x_0)\,\partial_x W_2(x,x_0)+\frac{1}{2}\sigma_x^2\,\partial_x^2 W_1(x,x_0)~, \\
  3W_3(x,x_0) &=& -(x-x_0)\,\partial_x W_3(x,x_0) + \frac{1}{2}\sigma_x^2\partial_x^2W_2(x,x_0)~,  \nonumber\\
                                                 &-& \frac{1}{2}\sigma_x^2(\partial_xW_1(x,x_0))^2 ~.
\end{eqnarray}
The differential equations above (\ref{makrieq}) are all first order, linear and inhomogeneous of the form
\begin{equation}
n W_n(x,x_0) = -(x-x_0) \partial_x W_n(x,x_0) + \Lambda_{n-1}(x,x_0)~,
\label{protomakri}
\end{equation}
where $\Lambda_{n-1}(x,x_0)$ is a function that is completely determined 
by the first $n-1$ relations. It can be readily verified 
by substitution and integration by parts that the solution of 
(\ref{protomakri}) reads 
\begin{equation}
W_n(x,x_0) = \int_0^1 \,d\xi \xi^{n-1} \Lambda_{n-1}(x_0+(x-x_0)\xi,x_0)~.
\end{equation}
This, for $n = 1,2,3$, after some manipulations, gives:
\begin{eqnarray}
	W_1(x,x_0) &=& \frac{1}{\Delta x} \int_{x_0}^x dz V_{\rm eff}(z)~, \label{expansion1} \\
	W_2(x,x_0) &=& \frac{ \sigma_x^2 } { 2\Delta x^2 } \left[ V_{\rm eff}(x)+V_{\rm eff}(x_0) - 2W_1(x,x_0) \right]~, \label{expansion2}  \\
	W_3(x,x_0) &=& -\frac{\sigma_x^2}{2\Delta x^4} 
	\left[ \Delta x \int_{x_0}^x dz V_{\rm eff}(z)^2 - \left(\int_{x_0}^x dz V_{\rm eff}(z)\right)^2\right]
   \nonumber \\
	&-& \frac{3 \sigma_x^2}{\Delta x^2} W_2(x,x_0) + \frac{\sigma_x^4}{4\Delta x^3}
	\left[\partial_x V_{\rm eff}(x)-\partial_x V_{\rm eff}(x_0)\right]~,
\label{expansion3}
\end{eqnarray}
where $\Delta x = x-x_0$, and,  for reasons that will be clearer in the next Section,
we have also introduced the `{\em effective potential}' as the following quantity with dimension
$time^{-1}$:
\begin{equation}
V_{\rm eff} (x) = \frac{1}{2\sigma_x^2} \mu_x(x)^2 + \frac{1}{2}\partial_x \mu_x(x)~.
\label{effpot}
\end{equation}
Note that, from  Eq.~(\ref{aitsa4}),  $V_{\rm eff}(x) = -2\lambda(x)$.
The first order correction can be rewritten as
\begin{equation}
W_1(x,x_0) = \frac{1}{\Delta t} \int_0^{\Delta t} dt \, V_{\rm eff} (x_0+ t \Delta x /\Delta t)~,
\end{equation} 
leading to the interpretation of this term as a time-average of the effective 
potential over the straight line, constant velocity ($\Delta x/\Delta t$)
trajectory between $x_0$ and $x$. Similarly, the leading term in $W_3(x,x_0)$ 
(i.e., the one proportional to the lowest power of the volatility) is proportional
to the variance of the effective potential over the same trajectory. 
Note that the corrections $W_n(x,x_0)$ are well defined in the limit $\Delta x \to 0$.
In particular, for $n=1,2,3$ it is not difficult to show that
\begin{eqnarray}
	\lim_{x\to x_0} W_1(x,x_0) &=&  V_{\rm eff}(x_0)~,  \\
	\lim_{x\to x_0} W_2(x,x_0) &=& \frac{\sigma_x^2}{12} \,\partial_x^2  V_{\rm eff}(x)~, \\
	\lim_{x\to x_0} W_3(x,x_0) &=& -\frac{\sigma_x^2}{24} \, ( \partial_x  V_{\rm eff}(x) )^2 +
	\frac{\sigma_x^4}{240} \, \partial_x^4  V_{\rm eff}(x) ~.
\label{limit}
\end{eqnarray}

Finally, the transition probability of the original diffusion (\ref{diffusionY}) is recovered by means of
the Jacobian transformation (\ref{aitsa5}).

The form of the trial transition probability represents the
main difference of the present approach to the one described in Section \ref{aitreview}  
Ref.~\cite{aitsahalia}, which is otherwise very similar in spirit. 
In fact, the latter expands in powers of $\Delta t$ the exponential $\exp{\left[-W(x,x_0,\Delta t)\right]}$ rather then just the exponent, as we do here, instead. As it will be shown explicitly in the following, the present choice 
gives rise to a distinct approximation scheme for $n>0$ providing generally a similar level of accuracy
but remarkably simpler mathematical expressions. In particular, all the calculations
related to the test cases presented in the following Sections have been easily obtained by hand, 
i.e., without the aid of symbolic calculation computer programs. This is because, 
by keeping the exponential form of the ansatz, one formulates a guess
which is closer to the exact one. The latter, can be expected 
to have an exponential form in order to satisfy the Chapman-Kolmogorov 
property of Markov processes \cite{shreve}.
In addition, the exponential choice of the ansatz 
automatically enforces the positive definiteness of the transition density which is not granted
in the approach of Ref.~\cite{aitsahalia}. In fact, as it will be shown explicitly in Sec.~\ref{app1},
in the limit of very small volatility ($\sigma_x \to 0$), when the effect of the noise disappears, and 
the
transition density converges to a Dirac's $\delta$ distribution, the expansion of Ref.~\cite{aitsahalia} 
breaks down as the transition probability becomes negative. On the contrary, the exponent 
expansion remains well defined and accurate also in this limit. Indeed, the first terms of the expansion in 
$\Delta t$ can be also derived through a small volatility expansion of the transition
density, as it will be discussed in Sec. \ref{pisec}.

Similarly to the approximation developed in Ref.~\cite{aitsahalia}, the exponent expansion 
has in general a finite convergence radius which is a decreasing function of the volatility. As it will be shown 
in the following, for the values of volatilities and $\Delta t$ relevant for financial applications the exponent expansion 
turns out to be very accurate even when truncated to the first few terms.
 
\subsubsection{Alternative derivation}
\label{alternative}
The term $W_0(x,x_0)$ in the exponent expansion
is somewhat different from the higher order terms. In fact,
it is defined by Eq.~(\ref{w0}) which is decoupled from the recursive
system (\ref{makrieq}). Indeed, it is possible to obtain 
the same result for the exponent expansion by expressing the 
transition density as
\begin{equation}
\rho_x(x,\Delta t|x_0) = e^{-W_0(x,x_0)} \Phi_{\rho_x}(x,\Delta t| x_0)~,
\end{equation}
and looking for an approximate expansion of the form (\ref{ansatz}) for $\Phi_{\rho_x}(x,\Delta t| x_0)$.
It is easy to show by direct substitution in the forward Kolmogorov 
equation (\ref{kolmogorov}) that $\Phi_{\rho_x}(x,\Delta t| x_0)$ is the solution of
\begin{equation}
{\cal H}_x \,\Phi_{\rho_x}(x,\Delta t| x_0) = - \partial_t \Phi_{\rho_x}(x,\Delta t| x_0)~, 
\label{schro} 
\end{equation}
where ${\cal H}_x$ is the ``Hamiltonian'' differential operator
\begin{equation}
{\cal H}_x = - \frac{\sigma_x^2}{2} \partial^2_x + V_{\rm eff}(x)~, 
\end{equation}
and $V_{\rm eff}(x)$ is the effective potential of Eq.~(\ref{effpot}).
As a result one can equivalently derive the exponent expansion 
by substituting in Eq.~(\ref{schro}) the following trial function
\begin{equation}
\Phi_{\rho_x}(x,\Delta t| x_0) = \frac{1}{\sqrt{2\pi\sigma_x^2\Delta t}} 
\exp {\left[ -\frac{(x-x_0)^2}{2\sigma_x^2 \Delta t}-\sum_{n=1}^\infty W_n(x,x_0) \Delta t^n\right]}~,
\label{trial2}
\end{equation}
which does not contain the term $W_0(x,x_0)$. 
This observation will be used in Section 3 to generalize the 
exponent expansion to the pricing kernel of financial derivatives.

\subsubsection{Exponent Expansion Coefficients in terms of the the Original Diffusion}

The derivations of the exponent expansion, and of the Hermite polynomial approximation
of Section \ref{aitreview}, both rely on the introduction of the auxiliary process (\ref{processX}),
and the integral transformation (\ref{inttransf}). As a result, the expressions obtained are easy to handle if 
the the volatility of the original process $\sigma_y(y)$ is such that both
the function $\gamma(y)$ (\ref{inttransf}) and its inverse $\gamma^{-1}(x)$ 
admit a closed-form expression. In fact, in this case the 
effective potential (\ref{effpot}) -- or the function $\lambda(z)$ (\ref{aitsa4}) -- 
has a closed-form expression, and the determination of the coefficients of the
expansion can be determined either analytically or through numerical quadrature.
This is the case for the examples considered in the following. 
However, these are very special cases. In fact, very few volatility
functions have a reciprocal for which a primitive exist. 
On the contrary, in many practical applications,  e.g. in local volatility 
models \cite{dupire}, the volatility is specified numerically through a fit to an
volatility function. For all these cases the application of the exponent 
expansion or the approach of Section \ref{aitreview} becomes cumbersome and 
computationally demanding, because it requires the numerical inversion of the
integral equation (\ref{inttransf}).

However, at a more careful analysis, it turns out that it is possible to circumvent this
difficulty, and to eliminate any dependence on the function  $\gamma^{-1}(x)$
in the expressions for the exponent expansion. This
makes its application straightforward for any diffusion process, 
irrespective of the analytic tractability of the specified volatility function. 
The first step to do this, is to note that, according to the Jacobian transformation (\ref{aitsa5}),
 $x$ and $x_0$ in Eq.~(\ref{ansatz}) need to be calculated for $\gamma(y)$, and $\gamma(y_0)$,
respectively. As a result, one can express the exponent expansion as
\begin{equation}
\rho_y(y,\Delta t|y_0) = \frac{1}{\sqrt{2\pi\Delta t\sigma_y(y)^2}}
\exp {\left[ -\frac{(\gamma(y)-\gamma(y_0))^2}{2\Delta t}- \sum_{n=0}^{\infty}\tilde W_n(y,y_0)\Delta t^n\right]}~.
\end{equation}
with the notation
\begin{equation}
\tilde G(y,y_0) = G(x,x_0)|_{x=\gamma(y), x=\gamma(y_0)},
\end{equation}
and taking, without any loss of generality $\sigma_x = 1$.
Now, using Eq.~(\ref{w0}), by means of the change  of variables $y = \gamma^{-1}(x)$ one can
express the zeroth order term as
\begin{equation}
\tilde W_0(y,y_0) = - \int_{y_0}^{y} \frac{dz}{\sigma_y(z)} \,\tilde \mu_x(z) \nonumber \\
\end{equation}
where, using Eq.~(\ref{drift}), 
\begin{equation}
 \tilde \mu_x(z) = \pm \left[ \frac{\mu_y(z)}{\sigma_y(z)} 
- \frac{1}{2}\frac{\partial \sigma_y}{\partial y}(z)\right]~.
\end{equation}
As anticipated, the expression for $W_0$ does not contain any reference to the
function $\gamma^{-1}(x)$ so that it can be calculated even if there is
no closed-form expression for $\gamma(y) = \int_{y_0}^y dz/\sigma_y(z)$ 
available, or if the latter is not analytically invertible.
Similarly, using Eqs.~(\ref{expansion1})-(\ref{expansion3}) one can easily find
\begin{eqnarray} \label{coeffnew1}
\tilde W_1(y,y_0) &=& \frac{1}{\Delta \tilde y} \int_{y_0}^y \frac{dz}{\sigma_y(z)} \tilde V_{\rm eff}(z)~, \\  \label{coeffnew2}
\tilde W_2(y,y_0) &=& \frac{1}{ 2 \Delta \tilde y^2 } \left[ \tilde V_{\rm eff}(y)+\tilde V_{\rm eff}(y_0) - 2 \tilde W_1(y,y_0) \right]~,  \\ \label{coeffnew3}
\tilde W_3(y,y_0) &=& -\frac{1}{2\Delta \tilde y^4}
        \left[ \Delta \tilde y \int_{y_0}^y \frac{dz}{\sigma_y(z)} \tilde V_{\rm eff}(z)^2 - 
        \left(\int_{y_0}^y \frac{dz}{\sigma_y(z)} \tilde V_{\rm eff}(z)\right)^2\right]
   \nonumber \\
        &-& \frac{3}{\Delta \tilde y^2} \tilde W_2(y,y_0) + \frac{\sigma_y(y)}{4\Delta \tilde y^3}
        \Big[\partial_y \tilde V_{\rm eff}(y)-\partial_y \tilde V_{\rm eff}(y)\Big]~,
\end{eqnarray}
where $\Delta \tilde y = \gamma(y) - \gamma(y_0)$, and the effective potential now reads
\begin{equation}\label{effpotnew}
\tilde V_{\rm eff} (y) = \frac{1}{2} \tilde \mu_x(y)^2 + \frac{1}{2} \sigma_y(y) \partial_y \tilde \mu_x(y)~.
\end{equation}
As a result, the exponent expansion can be easily applied for {\em any} specification of the volatility 
function $\sigma_y(y)$. Indeed, given any drift and volatility function one can immediately calculate 
 -- possibly numerically -- their derivatives and the function $\gamma(y)$. From these quantities, one
then easily calculates the effective potential (\ref{effpotnew}), and the coefficients of the exponent expansion (\ref{coeffnew1})-(\ref{coeffnew3}).

\begin{figure}[pt]
\centerline{\psfig{file=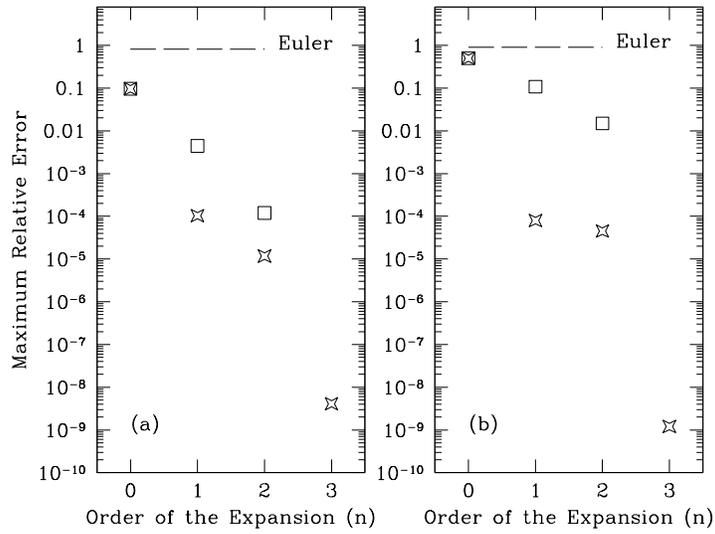,width=4.0in}}
\vspace*{-0.3cm}
\caption{Accuracy of the exponent expansion for the transition density of the Vasicek model (\protect\ref{vasicek}).
for $a=0.0717$, $b=0.261$ and $x_0=0.1$.  Panel (a):  comparison between the exponent expansion (stars) 
and the approximation of  Ref.~\protect\cite{aitsahalia} (squares) for $\sigma = 0.02237$ and $\Delta t = 0.5$.
At order zero the two schemes are identical. The uniform error of the Euler approximation is also
reported for comparison.  Panel (b): comparison between the exponent expansion (stars) and the approximation of  Ref.~\protect\cite{aitsahalia} (squares) in the regime of low volatility ($\sigma = 0.01$), for $\Delta t = 0.5$.}
\label{densvasicek1}
\end{figure}

\begin{figure}[pt]
\centerline{\psfig{file=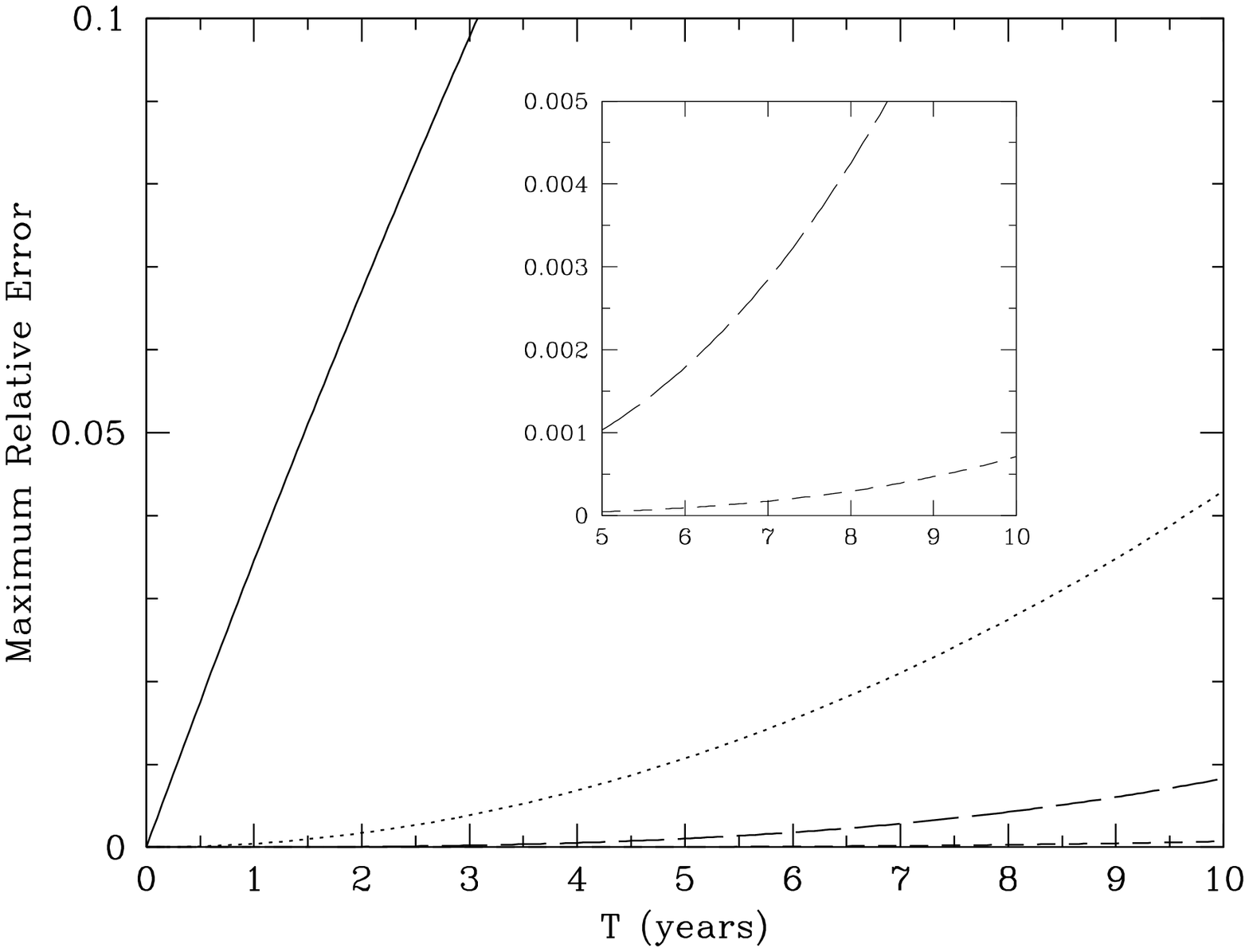,width=4.0in}}
\vspace*{-0.3cm}
\caption{Accuracy of the exponent expansion for the transition density of the Vasicek model (\protect\ref{vasicek}).
for $a=0.0717$, $b=0.261$ and $x_0=0.1$.  
Maximum relative error for $\sigma = 0.02237$ as a function of $\Delta t$: $n=0$ (continuous), $n=1$ (dotted), $n=2$ (long dashed) and $n=3$ (short dashed). The inset is an enlargement of the 5-10 years region.  }
\label{densvasicek2}
\end{figure}

\subsection{Examples}
\label{app1}
The application of the exponent expansion to a generic 
diffusion process of the form (\ref{diffusionY}) is rather 
straightforward and reduces to the calculation of one dimensional 
integrals.  In this Section, we illustrate this procedure 
for a few test cases, namely for the Vasicek \cite{vasicek}, the  
Cox-Ingersoll-Ross \cite{cir}, and the Constant Elasticity of Variance \cite{cev} 
diffusion processes. We will compare the results of the 
exponent expansion with the exact results available
in literature, and with the approach of Ref.\cite{aitsahalia}. 

\subsubsection{Vasicek diffusion}
\label{vasicekapp}
We first consider the Ornstein-Uhlenbeck diffusion 
proposed by Vasicek \cite{vasicek} as a model for the 
short-term interest rate:
\begin{equation}
d X_t= a ( b - X_t) dt + \sigma dW_t~, 
\label{vasicek}
\end{equation} 
where $a$, $b$, and $\sigma$ are positive constants representing
the mean-reversion level, the velocity to mean reversion, and
the volatility, respectively.  This model is integrable and 
the corresponding probability density function is Gaussian: 
\begin{equation}
\rho_{\rm ex}(x,\Delta t|x_0) = \frac{1}{(2\pi\bar\sigma^2)^{1/2}} 
\exp{\left[-\frac{[(x_0-a) e^ {-a \Delta t}-(x-a)]^2}{2\bar\sigma^2}\right]}~,
\label{vasicekexact}
\end{equation}
with
\begin{equation}
\bar \sigma = \sigma \sqrt{\frac{1-e^{-2a\Delta t}}{2 a}}~.
\label{vasiceksig}
\end{equation}

The exponent expansion of the Vasicek model can be easily derived 
using Eqs.~(\ref{w0}), and (\ref{expansion1}-\ref{expansion3}) with the
effective potential, Eq.~(\ref{effpot}),
\begin{equation}
V_{\rm eff} (x) = \frac{a^2(b-x)^2}{2\sigma^2}-\frac{a}{2}~,
\end{equation}
and gives,
\begin{eqnarray}
\rho_x(x,\Delta t|x_0) &=& \frac{1}{\sqrt{2\pi\sigma^2\Delta t}} 
\exp{\big [ -\frac{(x-x_0)^2}{2\sigma^2 \Delta t} - W_0(x,x_0) }  \nonumber \\ 
&-&  { W_1(x,x_0)\Delta t- W_2(x,x_0)\Delta t^2 - W_3(x,x_0)\Delta t^3 \big ]}~,
\end{eqnarray}
up to the third order in $\Delta t$ ($n=3$). Here
\begin{eqnarray}
W_0(x,x_0) &=& \frac{a (x-b)^2-a(x_0-b)^2}{2\sigma^2}~, \label{vas0} \\ 
W_1(x,x_0) &=& \frac{a^2}{6\sigma^2}((x-b)^2+(x_0-b)^2+(x-b)(x_0-b)) -\frac{a}{2}~, \label{vas1} \\
W_2(x,x_0) &=& \frac{\sigma^2}{2\Delta x^2}
\big[V_{\rm eff}(x)+V_{\rm eff}(x_0)-2W_1(x,x_0)\big]~, \label{vas2} \\
W_3(x,x_0) &=& -\frac{\sigma_x^2}{2\Delta x^3}\Big[ \frac{a^4}{20\sigma^4}[(x-b)^5-(x_0-b)^5]  \nonumber \\
           &+& \frac{a^2}{4}\Delta x -\frac{a^3}{6\sigma^2}[(x-b)^3-(x_0-b)^3]   \Big ] \nonumber \\
           &+& \frac{\sigma_x^2}{2\Delta x^2} (W_1(x,x_0))^2 - \frac{3\sigma_x^2}{\Delta x^2}W_2(x,x_0) 
           + \frac{a^2\sigma_x^2}{4\Delta x^2} ~, 
\label{vas3}
\end{eqnarray}
with $\Delta x =  x-x_0$.
It is interesting to note that the approximate transition probability obtained with 
the present approach reproduces  exactly the expansion of the exact transition 
density Eq.~(\ref{vasicekexact}) at the same order.  

On the other hand, the first two coefficients of 
the Hermite polynomials expansion (\ref{aitsa1}) as quoted in Ref.~\cite{aitsahalia}  read:
\begin{eqnarray}
c_1(x,x_0) & = & -\frac{1}{6\sigma^2} \Big[ a (3 b^2 a - 3(x+x_0) b a\sigma \nonumber  \\
 &+& (-3 + x^2 a+x x_0 a+x_0^2 a)\sigma^2 ) \Big]~,  \\
c_2(x,x_0) & = & \frac{1}{36 \sigma^4} \Big[a^2(9 b^4 a^2 - 18 x b^3 a^2 \sigma
+ 3 b^2 a (-6+5x^2 a)\sigma^2 \nonumber \\
&-& 6 x b a(-3 + x^2 a)\sigma^3+ (3-6 x^2 a+x^4 a^2)\sigma^4 \nonumber \\
&+& 2 a\sigma (-3 b + x\sigma)(3 b^2 a - 3x b a\sigma + (-3+x^2 a)\sigma^2)x_0 \nonumber \\
&+& 3 a\sigma^2(5 b^2 a-4x b a\sigma+(-2+x^2 a)\sigma^2)x_0^2 + 2 a^2\sigma^3 (-3 b+x\sigma)x_0^3 \nonumber \\
&+& a^2\sigma^4x_0^4)\Big]~.
\end{eqnarray}

The fast convergence of the approximation 
scheme is illustrated in Figs.~\ref{densvasicek1}\ref{densvasicek2}. Here the percentage error of the 
exponent expansion with respect to the exact result (\ref{vasicekexact}) is plotted 
for various $\Delta t$, and compared with the approach of Ref.~\cite{aitsahalia}. The parameter choice, also taken
from Ref.~\cite{aitsahalia} corresponds to a sensible parameterization for interest rate markets. We adopt one year
as unit of time, and we express the various parameters in this unit.  The Euler approximation
\begin{equation}
\rho_{E}(x,\Delta t | x_0) = \sqrt{\frac{1}{2 \pi \sigma_x^2\Delta t}} 
\exp{\left( -\frac{(x-x_0-\mu_x(x_0)\Delta t)^2} {2\sigma_x^2\Delta t}\right) }~,
\end{equation}
is also reported for comparison. 
The inclusion of each successive order allows one to increase dramatically the accuracy of the approximation
so that the third order expansion has basically a negligible error even for a sizable time step of order 
6 months. Remarkably, for the considered example, the third order expansion allows one to estimate the 
10 years transition probability with a relative error of less than 10 basis points (Fig.~\ref{densvasicek2}).
As illustrated in Fig.~\ref{densvasicek1}-(a), in the present case the approach of Ref.~\cite{aitsahalia} provides a slightly poorer level of accuracy for $n\leq 2$. In addition, it generally produces more complicated mathematical expressions. Furthermore, as shown in Fig.~\ref{densvasicek1}-(b), in the regime of small volatility  the exponent expansion still provides accurate results while the performance of the approach of Ref.~\cite{aitsahalia} degrades.
In fact, as anticipated, in the limit of small volatility ($\sigma_x \lesssim 0.5 \%$) 
the first order correction of the latter approach produces a negative transition probability signaling a break down of the scheme.

\subsubsection{Cox, Ingersoll and Ross diffusion}
\label{cirapp}

The Vasicek model is probably too easy of a test case 
as the associated transition probability is Gaussian. In fact, since 
the exponent expansion has a leading term which is Gaussian, the higher powers in $\Delta t$ 
just have to renormalize its average and variance in order to reproduce the exact result.
It is interesting therefore to test the accuracy of the exponent expansion
for a diffusion process that, while still integrable, generates a non-normal
transition density.  This is the case for the Feller's square root process \cite{feller} 
\begin{equation}
d Y_t = a (b - Y_t) + \sigma_y \sqrt{Y_t} dW_t
\label{cirdiff}
\end{equation}
which is the basis of Cox-Ingersoll-Ross  model 
for the instantaneous interest rate \cite{cir}.
The exact transition probability is given by \cite{cir}
\begin{equation}
\rho_{\rm ex}(y,\Delta t|y_0) = c e^{-(u+v)}\left(\frac{v}{u}\right)^{\frac{q}{2}}
I_q ( 2\sqrt{uv})~,
\end{equation}
where $c = 2a/[\sigma_y^2(1-\exp{(-a\Delta t)})]$, $q = 2 a b/\sigma_y^2 -1 \ge 0$, 
$ u = c y_0 \exp{(-a\Delta t})$, $v = c y$ and $I_q$ is the modified Bessel function
of the first kind of order $q$ \cite{abramovitz}.

As explained in Sec.\ref{expa}, since the volatility is not uniform, it 
is convenient to introduce the auxiliary process defined by Eq.~(\ref{inttransf}),
as $X_t = \gamma(Y_t) \equiv 2\sqrt{Y_t}/ \sigma_y$. The $X_t$ process follows 
Eq.~(\ref{processX}), with $\sigma_x=1$ and 
\begin{equation}
\mu_x(x) = \frac{\tilde{q}}{x}- \frac{a}{2} x~,
\end{equation}
and $\tilde{q}=q+1/2$.
In this case the effective potential reads:
\begin{equation}
V_{\rm eff} (x) = \frac{1}{2} \mu_x(x)^2 - \frac{\tilde{q}}{2x^2}-\frac{a}{4}~,
\end{equation}
and the first four  terms of the exponent expansion in Eq.~(\ref{w}) are:
\begin{eqnarray}
W_0(x,x_0) &=& - \tilde{q} \log{\frac{x}{x_0}} + \frac{a}{4} (x^2-x_0^2)  \label {cir0}\\ 
W_1(x,x_0) &=& \frac{1}{2\Delta x}
\Big[\mu_x(x) - \mu_x(x_0) 
- \tilde{q}^2 \left( \frac{1}{x} - \frac{1}{x_0}\right) \nonumber \\ 
&+& \frac{a^2}{12} \left(x^3-x_0^3\right) 
- a\tilde{q}\left(x-x_0\right)\Big]  \label {cir1} \\
W_2(x,x_0) & = &  \frac{1}{2\Delta x^2}
\big[V_{\rm eff}(x)+V_{\rm eff}(x_0)-2W_1(x,x_0)\big] \label{cir2} \\
W_3(x,x_0) &=& -\frac{1}{2\Delta x^2} 
	\left[  \frac{G(x)-G(x_0)}{\Delta x} - (W_1(x,x_0))^2\right] 
    \nonumber \\
	&-& \frac{3}{\Delta x^2} W_2(x,x_0) + \frac{1}{4\Delta x^3}
	\left[\partial_x V_{\rm eff}(x)-\partial_x V_{\rm eff}(x_0)\right]~,
\label {cir3}
\end{eqnarray}
where $\Delta x =  x-x_0$, $\partial_x V_{\rm eff}(z) = \tilde{q}(1-\tilde{q})/z^3+a^2 z/4$, and
\begin{equation}
G(z) = \frac{1}{5}\alpha^2 z^5 - \frac{1}{3}\frac{\beta^2}{z^3}+\gamma^2 z+2\alpha\beta z
- 2\frac{\beta\gamma}{z} + \frac{2}{3}\alpha\gamma z^3~,
\label{gz}
\end{equation}
and $\alpha = a^2/8$, $\beta = \tilde{q}(\tilde{q}-1)/2$, $\gamma = - a(\tilde{q}+1)/2$.
Finally, going back to the original process with Eq.~(\ref{aitsa5}), 
the transition probability reads:
\begin{equation}
\rho_y (y, \Delta t| y_0) = \rho_x (2\sqrt{y}/\sigma_y, \Delta t| \,2\sqrt{y_0}/\sigma_y)/\sigma_y \sqrt{y}~.  
\end{equation}

The coefficients of the expansion in Hermite polynomials Eq.~(\ref{aitsa1}) as quoted in Ref.~\cite{aitsahalia} 
read instead:
\begin{eqnarray}
c_1(x,x_0) & = & -\frac{1}{24 x x_0 \sigma^4} \Big[ 48 b^2 a^2 - 48 b a \sigma^2 + 9 \sigma^4 \nonumber \\
&+& x a^2 \sigma^2(-24b+x^2\sigma^2)x_0 + x^2 a^2 \sigma^4 x_0^2 + x a^2\sigma^4x_0^3 \Big]~, \\
c_2(x,x_0) & = & \frac{1}{576 x^2 x_0^2 \sigma^8} \Big[
9(256 b^4 a^4 - 512 b^3 a^3\sigma^2 + 224 b^2 a^2\sigma^4 + 32 b a \sigma^6 - 15 \sigma^8) \nonumber \\
&+& 6 x a^2 \sigma^2 (-24b + x^2\sigma^2)(16 b^2 a^2-16b a\sigma^2 + 3\sigma^4) x_0 \nonumber \\
&+& x^2 a^2\sigma^4 (672b^2 a^2 - 48 b a (2+x^2 a)\sigma^2 + (-6+x^4 a^2)\sigma^4) x_0^2 \nonumber \\
&+& 2 x a^2 \sigma^4 (48b^2 a^2 -24 b a (2+x^2 a)\sigma^2+(9+x^4 a^2)\sigma^4)x_0^3 \nonumber \\
&+& 3 x^2 a^4 \sigma^6 (-16 b + x^2\sigma^2)x_0^4 + 2 x^3 a^4\sigma^8 x_0^5 + x^2 a^4 \sigma^8 x_0^6 
 \Big]~.
\end{eqnarray}

\begin{figure}[pt]
\centerline{\psfig{file=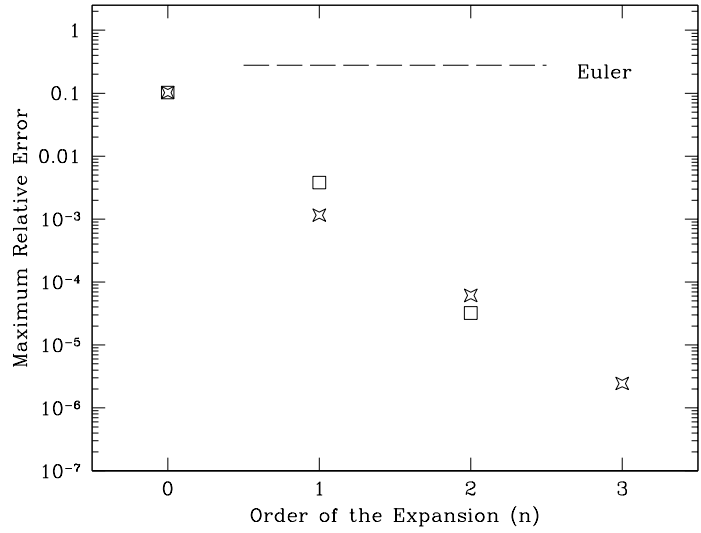,width=4.0in}}
\vspace*{-0.3cm}
\caption{Accuracy of the exponent expansion for the transition density of the Cox-Ingersoll-Ross model 
(\ref{cirdiff}), for $a=0.0721$, $b=0.219$, $\sigma = 0.06665$, and $x_0=0.06$.  Comparison between the exponent expansion (stars) 
and the approximation of  Ref.~\protect\cite{aitsahalia} (squares) for $\Delta t = 0.5$.
The uniform error of the Euler approximation is also
reported for comparison. 
}
\label{denscir1}
\end{figure}

\begin{figure}[pt]
\centerline{\psfig{file=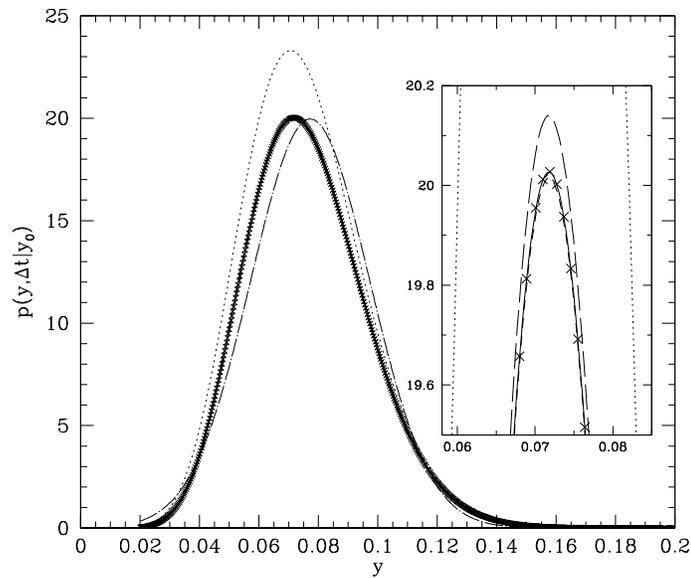,width=4.0in}}
\vspace*{-0.3cm}
\caption{Accuracy of the exponent expansion for the transition density of the Cox-Ingersoll-Ross model 
(\ref{cirdiff}), for $a=0.0721$, $b=0.219$, $\sigma = 0.06665$, and $x_0=0.06$.  
Probability density function  for $\Delta t = 1.5$ for different values of
$n$: $n=0$ (dotted), $n=1$ (long dashed), $n=2$ (short dashed), $n=3$ (continuous), 
Euler (dot-long dashed), exact (crosses). The inset is an enlargement of the region of the maximum. 
On this scale, the estimates for $n=2$ and $n=3$ still appear coincident. }
\label{denscir2}
\end{figure}

The accuracy of the exponent expansion in this case is illustrated in Figs.~\ref{denscir1} and \ref{denscir2}. 
Similarly to the case of the Vasicek diffusion, the exponent expansion is characterized by a 
remarkably fast convergence by including successive terms of the approximation so that
$n=3$ provides already a virtually exact representation of the transition density, for $\Delta t \simeq 1 \,yrs$.
In this case, the approach of Ref.~\cite{aitsahalia} performs slightly worse of the exponent expansion 
for $n=1$, and slightly better for $n=2$. However, also in this case the former breaks down
for small values of the volatility, generating unphysical transition densities. 

\subsubsection{Constant Elasticity of Variance diffusion}

As a last example we consider the Constant Elasticity of Variance model:
\begin{equation}
d Y_t = a(b-Y_t) dt + \sigma_y Y_t^p dW_t 
\end{equation}
Here we consider for brevity only the case $p>1$.
The transformation to a process with constant (unit)
variance is $X_t = \gamma(Y_t) = Y_t^{1-p}/\sigma_y(p-1)$
and gives, according to Eq.~(\ref{drift})
\begin{equation}
\mu_x (x) = \frac{c_1}{x}+ c_2 x + c_3 x^{\frac{p}{p-1}}~,
\end{equation}
with 
$c_1 = p/2(p-1)$, $c_2 = a(p-1)$, and $c_3 = - ab(p-1)^{p/(p-1)}\sigma_y^{1/(p-1)}$.  
In this case the effective potential reads
\begin{equation}
V_{\rm eff} (x) = \frac{1}{2} \mu_x(x)^2 - \frac{c_1}{2 x^2} + \frac{c_2}{2} + 
\frac{p}{2(p-1)} c_3 x^{1/(p-1)} ~,
\end{equation}
and the first three terms of the expansion are:
\begin{eqnarray}
W_0(x,x_0) &=& c_1\log{\frac{y_0}{y}} - \frac{a(p-1)}{2(2p-1)}
\Big[ (2p-1)(x^2-x_0^2)  \nonumber \\
&+& 2 b(p-1)^{\frac{p}{p-1}} \sigma_y^{\frac{1}{p-1}} 
\Big(x^{\frac{2p-1}{p-1}} - x_0^{\frac{2p-1}{p-1}}\Big )  \Big]~, \\ 
W_1(x,x_0) &=& \frac{1}{2\Delta x} \Big[ F(x)-F(x_0) \Big]~,\\
W_2(x,x_0) & = &  \frac{1}{2\Delta x^2}
\big[V_{\rm eff}(x)+V_{\rm eff}(x_0)-2W_1(x,x_0)\big]~, 
\end{eqnarray}
with 
\begin{eqnarray}
F(z) &=& -\frac{c_1^2}{z} + \frac{c_2^2}{3} z^3 + \frac{c_3^2(p-1)}{3p-1} z^{\frac{3p-1}{p-1}} \nonumber \\
 &+& 2 c_1c_2 z + \frac{2 c_1 c_3 (p-1)}{p}z^{\frac{p}{p-1}} 
 +  \frac{2 c_2 c_3(p-1)}{3p-2} z^{\frac{3p-2}{p-1}} + \mu_x(z)~. 
\end{eqnarray}

Similarly to the examples considered previously, also for the Constant Elasticity of Variance
model we find a very fast convergence of the exponent expansion for $\Delta t\simeq 1$, and a 
performance generally similar to the one of the approach of Ref.~\cite{aitsahalia}, for values
of the volatility large enough.

\section{PRICING KERNELS OF CONTINGENT CLAIMS}
\label{prickern}

\subsection{Path integral formulation} 
\label{pisec}
The exponent expansion can be generalized to obtain an approximation
of the pricing kernels of `standard' derivatives. 
This can be done by formulating the pricing problem within Feynman's path integral 
framework \cite{rosaclot1,rosaclot2}.  Here we indicate as `standard' any contingent claim written 
on the underlying, $Y_t$, whose value at time $t=0$, $V_0$, can be expressed as an expectation 
value of a functional of a certain type, namely
\begin{equation}
V_0(\Delta t, y_0) = E\Big [P(Y_{\Delta t}) F[Y_u] \Big| y_0 \Big ]~,
\label{st1}
\end{equation}
where
\begin{equation}
F[Y_u] = \exp{\left[ - \int_0^{\Delta t} du \, V_F[Y_u] \right]}~, 
\label{st2}
\end{equation}
and $P(Y_{\Delta t}) $ is a payout function.
Above $\Delta t$ is the time to expiry, and the expectation value 
is performed with respect to the probability measure defined 
by the diffusion for $Y_t$ that we assume of the form (\ref{diffusionY}).
European Vanilla options, zero coupon bonds, caplets, and floorets clearly belong to 
this family of contingent claims. In addition,  other path-dependent derivatives, 
like barrier or Asian options can be expressed in this form (see e.g., Refs.~\cite{baaquie,lin}). 

Similarly to the case of the transition probability, 
it is in general convenient to introduce an auxiliary diffusion 
with constant volatility of the form (\ref{processX})
by means of the integral transformations (\ref{inttransf}). 
Then, the expectation value in (\ref{st1}) can be transformed 
in an integral over the distribution generated by 
such auxiliary diffusion by means of the usual 
Jacobian transformation (\ref{aitsa5}). As a result, the value of the option
can be in general written as:
\begin{equation}
V_0(\Delta t, x_0) = E\Big [ P(X_{\Delta t}) F[X_u] \Big | x_0 \Big ] = \int_D dx P(x) K(x,\Delta t|x_0)~.
\label{integra}
\end{equation}
where $D$ is the domain of the auxiliary process  as defined by 
the relative stochastic differential equation (\ref{processX}), and $K(x, \Delta t|x_0)$ 
is the pricing kernel. The latter can be expressed in terms of 
a path integral as it follows\cite{rosaclot1,rosaclot2}
\begin{equation}
K(x,\Delta t|x_0) = e^{-W_0 (x,x_0)} \Phi(x,\Delta t| x_0) 
\label{kernel}
\end{equation}
with 
\begin{equation}
\Phi(x,\Delta t| x_0) = \int_{x(0)=x_0}^{x(\Delta t)=x} \, {\cal D}[x(u)] 
\exp{\left[ -\int_{0}^{\Delta t} du \, L_{\rm eff}[x(u),\dot x(u)] \right]}.
\label{pi}
\end{equation}
where $W_0$ is given by Eq.~(\ref{w0}) and
the functional $L_{\rm eff}[x(u),\dot x(u)]$ is the effective 
{\em Euclidean Lagrangian} 
\begin{equation}
L_{\rm eff}[x(u),\dot x(u)] =  \frac{1}{2\sigma_x^2} \dot x(u)^2 + V_{\rm eff}(x) 
\end{equation} 
($\dot x(u) \equiv dx(u)/du$) with the {\rm effective potential}, $V_{\rm eff}(x)$, defined as:
\begin{equation}
V_{\rm eff} (x) = \frac{1}{2\sigma_x^2} \mu_x(x)^2 + \frac{1}{2}\partial_x \mu_x(x) + V_F(x)~.
\label{effpot2}
\end{equation}
Finally, the measure ${\cal D}[x(u)]$ is defined by discretizing 
each path $x(u)$ connecting $x(0)=x_0$ and $x(T)=x$. This can be done by 
dividing the time interval $[0,T]$ into $P$ intervals so that 
$x_n = x(u_n)$ ($u_n = nT/P$ with $n=0,...,P$),  and by integrating the 
internal $P-1$ variables $x_n$ over the domain $D$. The path integral 
$\int {\cal D}[x(u)]$ is then obtained as the limit for 
$P\to \infty$ of  this procedure, namely
\begin{equation}
\int {\cal D}[x(u)] \equiv  \lim_{P\to\infty} (2\pi \sigma_x^2 \Delta t)^{-P/2} \prod_{n=1}^{P-1} \int_D \, d x_n ~.  
\end{equation}

It is well known form the physical sciences that the path integral 
$\Phi(x,\Delta t| r_0)$ satisfies the partial differential equation 
Eq.~(\ref{schro}) \cite{pireference1,pireference2}. Note that this is consistent with the fact that, 
by definition,  for $V_F(x) \equiv 0$ the pricing kernel coincides 
with the transition density of the underlying diffusion process for  $X_t$. 
In particular, as observed in Sec.~\ref{alternative}, one can use Eq.~(\ref{schro}) 
to derive the exponent expansion for $\Phi(x,\Delta x| r_0)$ using the trial form
(\ref{trial2}). As a result, the same expressions Eqs.(\ref{expansion1}-\ref{expansion3}) 
derived for the transition  density hold true also 
for the pricing kernel, provided that the effective potential (\ref{effpot2}) 
replaces the one in Eq.~(\ref{effpot}). 

\subsection{Correspondence with Quantum Mechanics}
\label{qm}

It is interesting to note that the path integral formulation
of the pricing kernel (\ref{pi}) is mathematically 
equivalent to the quantum mechanical description of the 
thermodynamic properties of an ideal gas of particles moving 
in the potential $\hbar V_{\rm eff}(x)$ ($\hbar$ is the reduced Planck's constant giving
the correct energy dimensions). The complete correspondence is 
obtained by identifying $\sigma_x^2 \to \hbar/m$ and $\Delta t \to \hbar/k_B T$
where $m$ is the mass of the particle, $T$ is the temperature, and $k_B$ is the Boltzmann constant. 
With this prescription, $\Phi(x,\Delta t| x_0)$ becomes the so-called 
single particle density matrix, and the results of Makri and Miller \cite{makri} can
be readily recovered. In addition, it is straightforward to show using Eqs.~(\ref{limit}) 
that the exponent expansion of its diagonal elements, $\Phi(x_0,\Delta t| x_0)$,  are consistent 
with the so called  Wigner expansion for the high-temperature limit.
Finally, performing the analytic continuation known as Wick 
rotation $\Delta t \to i \hbar t$ allows one to obtain the single-particle quantum 
propagator.
In this case (\ref{schro}) becomes the celebrated Schr\"odinger 
equation. 

This correspondence provides an alternative justification of  the choice of the exponential ansatz
in Eq.~(\ref{ansatz}). Indeed, this is the form in which can be expressed in general 
the quantum mechanical propagator or the single particle density matrix \cite{pireference1,pireference2}. 
Furthermore, it has been shown for the 
quantum problem \cite{makri2} that the exponent expansion up to third order in $\Delta t$ and 
first order in $\hbar/m$ can be derived starting from the short time semiclassical 
propagator obtained through a saddle point analysis of the 
limit $\hbar/m\to 0$ \cite{schulman}. Indeed, it can be shown that the second order correction 
$W_2(x,x_0)$, Eq.~(\ref{expansion2}),  is the so-called van-Vleck determinant of the
saddle point expansion.  This explains why the accuracy
of the present scheme is preserved in the corresponding regime of low volatility, as it was
anticipated in Sec.\ref{expa}, and illustrated in Sec.\ref{app1}. 

\begin{figure}[pt]
\vspace*{-1.5cm}
\centerline{\psfig{file=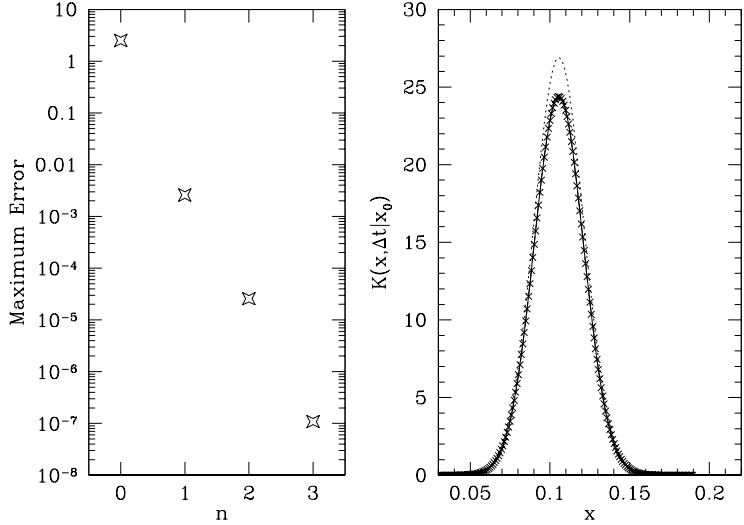,width=4.0in}}
\vspace*{-1cm}
\caption{Accuracy of the exponent expansion for the pricing kernel Eq.~(\ref{kernel}) 
of the Vasicek model (\ref{vasicek}), for $a=0.0717$, $b=0.261$, and $x_0=0.1$. 
Symbols as in Fig.~\ref{denscir2}. The panel on the left shows the maximum absolute error as a 
function of the order of the expansion $n$. }
\label{prop}
\end{figure}

\subsection{An example: Zero Coupon Bond}    
\label{app2}
In this Section we illustrate the prescriptions outlined above
by applying the exponent expansion to the calculation of a 
zero coupon bond within the Vasicek and Cox-Ingersoll-Ross models.
The zero coupon bond is a financial instrument that provides at time $t=\Delta t$
a payout of one unit of a certain notional. Its value at time $t=0$ can
be expressed therefore as
\begin{equation} 
P(0,\Delta t) = E\Big[ \exp{ - \int_0^{\Delta t} du \, X_u } \Big | r_0 \Big]
\end{equation}
which is of the standard form given by Eqs.~(\ref{st1}) and (\ref{st2}), with 
$V_F[X_u] = X_u$, and $P[X_{\Delta t}] \equiv 1$.  

As a result, the exponent expansion for the kernel Eq.~(\ref{kernel})
can be easily derived giving for the Vasicek model
\begin{eqnarray}
W_1(x,x_0) &=& W_1^0(x,x_0) + \frac{x+x_0}{2}\\
W_2(x,x_0) &=& W_2^0(x,x_0) \\
W_3(x,x_0) &=& W_3^0(x,x_0) - \frac{\sigma_x^2 +  2 a^2 (x-b) }{24}    
\end{eqnarray}
where $W_i^0(x,x_0)$ are the expressions obtained for the transition probability 
Eqs.~(\ref{vas0})-(\ref{vas3}) of Sec.~\ref{vasicekapp}. For the Cox-Ingersoll-Ross model 
instead we get:
\begin{eqnarray}
W_1(x,x_0) &=& W_1^0(x,x_0) + \frac{\sigma^2}{12}(x^2+x_0^2+xx_0) \\
W_2(x,x_0) &=& W_2^0(x,x_0) - \frac{\sigma^2}{24} 
\end{eqnarray}
with $W_i^0(x,x_0)$ given by Eqs.~(\ref{cir0})-(\ref{cir2}), and
$W_3^0(x,x_0)$ related to the previous quantities as in Eq.~(\ref{cir3})
with
\begin{equation}
G(z)=G^0(z) + \frac{\sigma_y^2}{2}\left[ (\alpha +\frac{\sigma_y^2}{8})\frac{z^5}{5}
+\frac{\gamma}{3}z^3+ \beta z \right]~, 
\end{equation}
$G^0(z)$ as in Eq.~(\ref{gz}), and $\partial_z V_{\rm eff}(z) = \partial_z V^0_{\rm eff}(z) + \sigma_y^2 z /2$.
The exponent expansion for the pricing kernel can be compared with the
exact results that can be shown to read for the Vasicek model 
(\ref{vasicek})
\begin{eqnarray}
K_{\rm ex}(x,\Delta t|,x_0) &=& \frac{\exp{\left[(x-x_0)/a-\Delta t(b-\sigma^2/2a^2 )\right]}}{(2\pi\bar\sigma^2)^{1/2}} 
\nonumber \\ 
&&\exp{\left[-\frac{\left[\left (x_0-b+\sigma^2/a^2 \right) e^ {-a \Delta t}-(y-b+\sigma^2/a^2 )\right]^2}
{2\bar\sigma^2}\right]}~,
\end{eqnarray}
with $\bar\sigma$ given by Eq.~(\ref{vasiceksig}), and
\begin{eqnarray}
&&K_{\rm ex}(x,\Delta t|,x_0) = \frac{2}{x}
\exp{\left[\frac{-\frac{a}{4}(x^2-x_0^2)+(2ab-\frac{\sigma^2}{2})\log{\frac{x}{x_0}}}{\sigma^2}\right]} 
\frac{\gamma \sqrt{xx_0}\,e^{a^2b\Delta t/\sigma^2}}{2\sigma^2\sinh\left[\gamma\Delta t /2\right]} 
\nonumber \\  &&
\exp{\left[-\frac{\gamma}{4\sigma^2}(x^2+x_0^2)\coth{\left[\gamma\Delta t/2\right]}\right]}
I_q\left(\frac{xx_0\gamma}{2\sigma^2\sinh{\left[\gamma\Delta t/2\right]}}\right)
\end{eqnarray}
with $\gamma =\sqrt{a^2+2\sigma^2}$, for the Cox-Ingersoll-Ross one.
As illustrated in Fig.\ref{prop}, similarly to the case of the
transition probability, the exponent expansion provides a remarkably good, and
fast converging approximation of the exact pricing kernel for financially sensible
parameterizations, and for a sizable value of the time step $\Delta t$.

\begin{figure}[pt]
\vspace{-1.6cm}
\centerline{\psfig{file=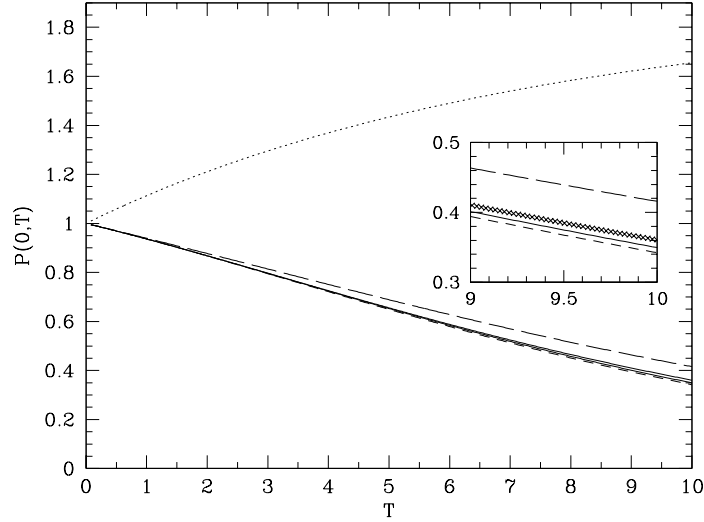,width=4.0in}}
\vspace*{-1cm}
\caption{Zero Coupon Bond for the Cox-Ingersoll-Ross model. Parameters and symbols as in Fig.\protect\ref{denscir2}.}
\label{zerof}
\end{figure}

Finally, the zero coupon bond can be  obtained by numerical integration 
of the pricing kernel according to Eq.~(\ref{integra}). The corresponding 
results are shown in Fig.~\ref{zerof} confirming once more the quality of 
the approximation. In a similar fashion, one can obtain systematic approximations 
for caplets, floorets, and other simple interest rate derivatives whose value depends 
only on the value of the instantaneous short rate at time $t = \Delta t$. It is also possible 
to generalize this approach to path dependent contingent claims, like Asian options.

\section{EXTENDING THE TIME STEP: PATH INTEGRAL MONTE CARLO METHODS}
\label{montecarlo}

For an extended time interval $T$, the calculation of the transition density 
or, more in general, of the pricing kernel (\ref{kernel}) can be 
performed by discretizing the path integral (\ref{pi}), and taking the limit of large 
number of time slices ($P\to \infty$) according to the standard Trotter product formula:
\begin{eqnarray}
& &K(x,\Delta t|x_0) \simeq  (2\pi \sigma_x^2 \Delta t)^{-P/2} e^{-W_0 (x,x_0)}
\prod_{n=1}^{P-1} \int_D \, d x_n \nonumber \\
& \times &\exp{\left[-\frac{1}{2\sigma_x^2\Delta t}\sum_{k=1}^{k=P}(x_k-x_{k-1})^2 - 
\frac{\Delta t}{2} \sum_{k=1}^{k=P} \left( V_{\rm eff}(x_k)+V_{\rm eff}(x_{k-1})\right)  \right]} 
\label{finiteT}
\end{eqnarray}
with $\Delta t = T/P$, $x_P=x$, and the effective potential given by Eq.~(\ref{effpot2}). 
It is worth remarking that, for the case of the transition density, 
by interpreting the latter equation as the Chapman-Kolmogorov property of Markov processes \cite{shreve} one 
obtains the following approximation of the short-time propagator
\begin{equation}
K_{\rm Trotter}(x,\Delta t|x_0) = \frac{e^{-W_0 (x,x_0)}}{\sqrt{2\pi\sigma_x^2\Delta t}} 
\exp {\left[ -\frac{(x-x_0)^2}{2\sigma_x^2 \Delta t}- \frac{\Delta t}{2} 
\Big( V_{\rm eff}(x)+V_{\rm eff}(x_{0})\Big) \right]}~.
\end{equation}
However, in contrast to the $n=1$ exponent expansion, the latter expression is 
not strictly correct up to order $\Delta t$, and only in the limit $P\to \infty$ 
the difference becomes negligible. 

In general, to obtain an accurate result for the pricing kernel 
for an extended time period $T$ one has to increase the number of time slices, 
or Trotter number $P$, until convergence is achieved. By replacing the Trotter formula 
with the improved short-time kernel obtained through the exponent expansion (\ref{ansatz}) 
one achieves a faster convergence with the Trotter number, thus significantly 
reducing the computational burden. In this case the finite-time expression of the 
pricing kernel reads
\begin{eqnarray}
& &K(x,T|x_0) \simeq  (2\pi \sigma_x^2 \Delta t)^{-P/2} 
\prod_{n=1}^{P-1} \int_D \, d x_n \nonumber \\
& \times &\exp{\left[-\frac{1}{2\sigma_x^2\Delta t}\sum_{k=1}^{P} (x_k-x_{k-1})^2 - 
\sum_{k=1}^{P} W(x_k,x_{k-1},\Delta t)  \right]} 
\label{finiteT2}
\end{eqnarray}
with $W(x,x_0,\Delta t)$ given by Eq.~(\ref{w}).

Equation (\ref{finiteT2}) allows one to obtain the transition density
or the pricing kernel for a standard derivative through the calculation of a multidimensional 
integral over the variables $x_1, \dots, x_{P-1}$. The latter integration is ideally 
suited for Monte Carlo methods either in real, or in Fourier space \cite{quasifourier}, the specific choice
depending on the particular problem at hand. 
In addition, importance-sampling schemes, e.g., by means of the Metropolis algorithm \cite{metropolis}, can be easily applied in order
to reduce the computation time. However, in order not to introduce a systematic bias in the result
a particular attention has to be paid in order to sample accurately the configuration space.  

The most straightforward way to perform a Monte Carlo quadrature of Eq.~(\ref{finiteT2}), 
is to realize that a simple Markov chain ${\bf x} = (x_1,\dots,x_{P-1})$
\begin{equation}
x_n = x_{n-1} + \sigma_x \sqrt{\Delta t} \, Z_n~, 
\label{markov}
\end{equation}
with $Z_n$, sampled from a standard normal distribution, generates an ensemble of walkers distributed according to
\begin{equation}
\rho(x_1,\dots,x_{P-1}|x_0) = (2\pi \sigma_x^2 \Delta t)^{-(P-1)/2} \exp{\left[ -\frac{1}{2\sigma_x^2\Delta t}\sum_{k=1}^{P-1}(x_k-x_{k-1})^2\right] }~.
\end{equation}
As a result, the pricing kernel (\ref{finiteT2}) can be obtained as the average over 
the random paths generated according to Eq.~(\ref{markov}) of the following estimator:
\begin{equation}
{\cal O}({\bf x}, x_P =x ) =  (2 \pi \sigma_x^2 \Delta t)^{-1/2} \exp{ \left[-\frac{1}{2\sigma_x^2\Delta t} (x-x_{P-1})^2 - \sum_{k=1}^{P} W(x_k,x_{k-1})  \right]}~.
\end{equation}
A remarkable property of the path integral approach is that $K(x,\Delta t|x_0)$
for any final point $x$ can be evaluated with a single Monte Carlo 
simulation by appropriately reweighting the accumulated estimator. In fact the distribution
of walkers $p(x_1,\dots,x_{P-1}|x_0)$ is independent of the final point $x_P$ so that 
$K(x^\prime,\Delta t|x_0)$ can be calculated by averaging ${\cal O}({\bf x}, x_P = x^\prime)$. In addition,
the latter quantity can be efficiently obtained by means of the following reweighting procedure
\begin{equation}
{\cal O}({\bf x}, x_P = x^\prime) = {\cal O} ({\bf x}, x_P =x ) \frac{{\cal W}(x^\prime,{\bf x})}{ {\cal W}(x,{\bf x})}
\end{equation}
with
\begin{equation}
{\cal W}(x,{\bf x}) = \exp{\left[ -\frac{1}{2\sigma_x^2\Delta t} (x-x_{P-1})^2 -W(x,x_{P-1}) \right]}~.
\end{equation}

Expectation values  of the form (\ref{integra}) on a time horizon $T$ can be calculated 
by integrating over the final variable giving:
\begin{eqnarray}
V_0(T , x_0) &=&  \int_D dx P(x) K(x,T|x_0)  \simeq
(2\pi \sigma_x^2 \Delta t)^{-P/2}  \prod_{n=1}^{P} \int_D \, d x_n  P(x_P) \nonumber \\ & & \exp{\left[-\frac{1}{2\sigma_x^2\Delta t}\sum_{k=1}^{P} (x_k-x_{k-1})^2 - 
\sum_{k=1}^{P} W(x_k,x_{k-1})  \right]}~. 
\end{eqnarray}  
This can be simulated by extending the Markov chain (\ref{markov}) up to step $x = x_P$, and
accumulating the estimator
\begin{equation}
{\cal P}({\bf x}, x_P = x^\prime) =  P(x) \exp{\left[- \sum_{k=1}^{P} W(x_k,x_{k-1})  \right]}~.
\end{equation}
Remarkably, within the path integral approach, the sensitivities of such
expectation values with respect to the model parameters (the so-called Greeks) 
can be computed in the same Monte Carlo simulation, thus
avoiding any numerical differentiation.
Indeed, indicating with $\theta$ a generic parameter, 
under quite general conditions \cite{glassermann}, one has
\begin{equation} 
\partial_\theta V_0(T,x_0,\theta)= \int_D dx \left[ K_\theta(x,T|x_0) \partial_\theta P(x,\theta) + P(x,\theta)\partial_\theta {K_\theta(x,T|x_0)}\right]~.
\end{equation} 
 As a result the sensitivity 
$\partial_\theta V_0(T,x_0,\theta)$ can be calculated by means of the estimator:
\begin{equation}
{\cal G}({\bf x}, x_P = x^\prime) =   \exp{\left[- \sum_{k=1}^{P} W(x_k,x_{k-1})\right]} 
\left (\partial_\theta P + P \partial_\theta \log{K_\theta} \right)~.
\end{equation}
Higher order sensitivities can be obtained in a similar fashion.

\begin{figure}[pt]
\centerline{\psfig{file=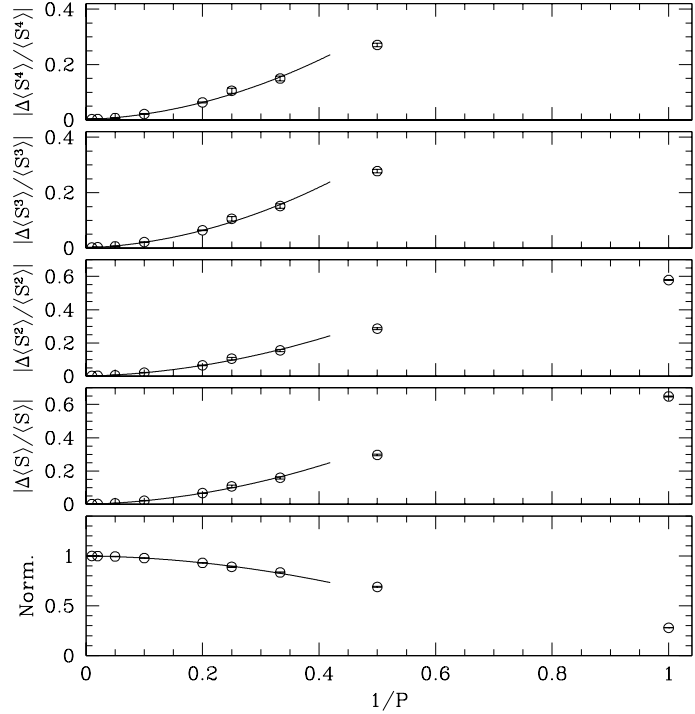,width=4.0in}}
\caption{Convergence with the Trotter number $P$ 
of the normalization, and of the first 4 moments 
of the 40 years transition density for the Cox-Ingerssol-Ross model. 
Parameters as in Fig.\protect\ref{denscir1}. In the first top panels the 
relative error with respect to the exact result is reported. Lines are quadratic fits.}
\label{pimc}
\end{figure}

The convergence with the Trotter number $P$ of  the path integral Monte Carlo estimates 
is illustrated in Fig.~\ref{pimc} for the calculation of the first five moments 
of the $T=40\,yrs$ transition probability of the Cox-Ingersoll-Ross model (\ref{cirdiff}). 
The  finite $P$ estimates converge very rapidly with $1/P$. In particular, for the case considered, 
$P=20$ already provides estimates in agreement with the exact result 
within statistical uncertainties. In general, as also
shown in the figure, a convenient indicator of the convergence is the zeroth-moment or normalization of
the distribution. The calculation of this quantity allows in general to assess the level 
of convergence without performing a complete scaling with $P$, thus saving computational time.

\section{EXTENDING THE TIME STEP: GREEN'S FUNCTION BACKWARD INDUCTION}

The {\em Green's function backward induction} is a deterministic approach 
alternative to partial differential equation methods,
introduced in financial context by Rosa-Clot and Taddei \cite{rosaclot2}.
This is based on the path integral representation
of standard contingent claims (\ref{st1}), that we rewrite here for convenience as
\begin{equation}
E\left[G_T\left(Y_u\right)\mid Y_0 \right]~.
\label{expstand}
\end{equation}
Here $G_T$ is the payout functional dependent on the realization of the path of the underlying up
to expiry $T$, and $Y_t$ is the general process following the diffusion (\ref{diffusionY}).
As partial differential methods, the Green's function backward induction is efficient for
low-dimensional problems but allows one to include naturally early exercise features in
the contingent claim, i.e., to treat expectation values of the form
\begin{equation}
\max_{\tau} E\left[G_\tau\left(Y_u\right)\right]
\end{equation}
where $0 \le \tau \le T$ is the {\em stopping time} of the process \cite{shreve}.

For standard contingent claims,  Eqs.~(\ref{st1}) and (\ref{st2}),
the functional $G_T[Y_u]$ can be discretized in the form,
\begin{equation}
G_T[Y_u]\simeq \prod_{i=0}^{N} g^{(i)}(Y_i).
\end{equation}
and the conditional expectation value (\ref{expstand}) is given by
\begin{equation}
E[G_T[Y_u]\mid Y_0]\simeq \int\!...\!
\int \prod_{i=1}^{N} dY_i\;
g^{(N)}(Y_{N})\; \prod_{j=1}^{N} {\tilde\rho}(Y_j,t_j \mid Y_{j-1},t_{j-1}),
\label{e:condapp}
\end{equation}
where the function, ${\tilde\rho}$, is
\begin{equation}
{\tilde\rho}(Y_i,t_i\mid Y_{i-1},t_{i-1}) =
\rho_y(Y_i,t_i - t_{i-1}\mid Y_{i-1})\; g^{(i-1)}(Y_{i-1})~,
\end{equation}
and $\rho_y(Y_i,t_i - t_{i-1}\mid Y_{i-1})$ is the transition
probability associated with the process (\ref{diffusionY}).  
This is the Green's function of the partial differential equation associated
with the calculation of the expectation value (\ref{expstand}), hence
the name of the method.

Let us now consider a single integration
\begin{equation}
\int dY_i\; {\tilde\rho}(Y_{i+1},t_{i+1}\mid Y_i,t_i)\;
{\tilde\rho}(Y_i,t_i\mid Y_{i-1},t_{i-1})\;.
\label{e:single}
\end{equation}
If we approximate this integral by using a numerical quadrature
rule, we obtain the following algebraic relation
\begin{equation}
\sum_{\gamma=1}^M\;
{\tilde\rho}^{~(i)}_{\alpha\gamma}\;\;{\tilde\rho}^{~(i-1)}_{\gamma\beta}\;
w_\gamma\;,
\end{equation}
where the matrices, ${\tilde\rho}^{~(i)}$, are defined by
\begin{equation}
{\tilde\rho}^{~(i)}_{\alpha\beta}=
{\tilde\rho}(z_\alpha,t_{i+1}\mid z_\beta,t_i),
\end{equation}
the quantities, $w_\alpha$, and $z_\alpha$, are the weights and the grid
points, respectively, associated with the integration rule, and
$\alpha,\beta,\gamma=1,\ldots,M$.
In conclusion, the expression (\ref{e:condapp}) can be written as
\begin{equation}
E[G_T[Y(\tau)]\mid Y_0=z_\alpha]
\simeq \sum_{\gamma_1,\ldots,\gamma_N=1}^M\;
G^{(0)}_{\gamma_1\alpha}\; G^{(1)}_{\gamma_2\gamma_1}\; \ldots
G^{(N-2)}_{\gamma_{N-1}\gamma_{N-2}}
G^{(N-1)}_{\gamma_N\gamma_{N-1}}\;g^{(N)}_{\gamma_N}\;\;\;,
\label{e:prodg}
\end{equation}
where $G^{(i)}_{\alpha\beta}=
w_{\alpha}\,{\tilde\rho}^{(i)}_{\alpha\beta}$, and $g^{(N)}_{\gamma_N}=
g^{(N)}(z_{\gamma_N})$.
Therefore, we have reduced the evaluation of the expectation
value of a functional to the product of $N$ matrices with dimension $M$.
By starting the calculation from the
right (backward induction), we need to memorize just linear arrays,
while the matrix elements, $G^{(i)}_{\alpha\beta}$, can be computed step by
step. In practice, one can follow the following algorithm:

\begin{itemize}
\item[i.] Set $u_\alpha=g^{(N)}_\alpha$ ($\alpha=1,\ldots,M$), and $i=N-1$.

\item[ii.] Set
$v_\alpha=\displaystyle{\sum_{\beta=1}^M}\; u_\beta\,G^{(i)}_{\beta\alpha}$
($\alpha=1,\ldots,M$).

\item[iii.] If $i>0$ then set $u_\alpha=v_\alpha$ ($\alpha=1,\ldots,M$),
$i=i-1$, and go to ii.
\end{itemize}
Here the arrays, $u_\alpha$, and $v_\alpha$, are two working vectors.

The inclusion of early exercise features is completely straightforward
and simply involves the following steps:
\begin{itemize}
\item[i.] Set $u_\alpha=g^{(N)}_\alpha$ ($\alpha=1,\ldots,M$), and $i=N-1$.

\item[ii.] Set $w^{(i)}_\alpha=f(z_\alpha,t_i)$\/, and
$v_\alpha=
\max\left(\displaystyle{\;\sum_{\beta=1}^M}\;
u_\beta\,G^{(i)}_{\beta\alpha}\;,\;\;
w^{(i)}_\alpha\;\right)$

($\alpha=1,\ldots,M$).

\item[iii.] If $i>0$ then set $u_\alpha=v_\alpha$ ($\alpha=1,\ldots,M$),
$i=i-1$, and go to ii.
\end{itemize}
The only difference is in the point ii, where we have now a test operation,
and the function $f(z_\alpha,t_i)$ is the exercise value of the option.

The exponent expansion can significatively speed up the Green's function
backward induction presented above. In fact, as illustrated in the previous Sections,
it allows one obtain reliable approximations of 
${\tilde\rho}(Y_i,t_i\mid Y_{i-1},t_{i-1})\;$ for a time step $(t_i-t_{i-1})$ much larger than those that 
can be safely used with a simple Euler discretization. As a result, one can reduce
the discretization bias to acceptable levels with a much smaller number of
intermediate time steps $N$.
The application of this procedure to local volatility models is currently in progress
and will be presented elsewhere \cite{lucaump}.

\section{CONCLUSIONS}
\label{conclusion}

Closed-form approximations of non-liner  diffusions 
are of primary importance in a variety of fields of quantitative Finance.
In Econometrics, they are crucial for an efficient maximum likelihood estimation 
of the parameters of continuous time processes.
In derivative pricing,  they allow one to develop effective
approximation schemes or to improve the efficiency of 
numerical approaches. 

In this paper we have presented an effective method to produce 
a family of closed-form approximations of the transition probability 
of a general diffusion process. 
Such approximation, dubbed {\em exponent expansion}, 
is based on an exponential ansatz of the transition probability for a finite time interval $\Delta t$,
and a series expansion of the deviation of its logarithm from that of a Gaussian
distribution. Through this procedure the transition probability is
obtained as a power series in $\Delta t$ which becomes asymptotically exact
if an increasing number of terms is included, and provides remarkably accurate
results even when truncated to the first few (say 3) terms. 
This approach  can be easily implemented, and involves 
the calculation of simple one dimensional integrals. 
In addition,  it applies to a very wide class of volatility functions, 
even to those that do not have a simple analytic expression but they are
specified through a numerical interpolation.

We have shown that the exponent expansion produces very accurate results 
for integrable diffusions of financial interest, like the Vasicek and the Cox-Ingersoll-Ross models. 
In particular, we have compared our results with those obtained with
the state-of-the-art approximation of discretely
sampled diffusions \cite{aitsahalia}, that shares a similar rationale.
We find that the exponent expansion provides a similar accuracy
in most of the cases but it is more stable in the low-volatility regimes.
Furthermore the implementation of the exponent expansion turns out to be
simpler.

By introducing a path integral framework we have generalized the exponent expansion to the calculation of the
pricing kernels of financial derivatives, and we have shown how to obtain 
accurate  approximations for the price of simple contingent 
claims. We have also shown how the exponent expansion can be naturally used
to increase the efficiency of both deterministic and stochastic numerical simulations. 
A systematic  study of the efficiency of this approach for the pricing of exotic derivatives, 
and the calibration of local and stochastic volatility models will be the object of
future investigations.

{\bf Acknowledgments} It is a pleasure to thank Gabriele Cipriani for sparking my interest 
in the path integral approach to option pricing, and for continuous stimulating discussions.
I would also like to thank Marco Rosa-Clot for kindly sharing with me a 
series of unpublished results.


\begin{thebibliography}{00}

\bibitem{bachelier} L. Bachelier,  Theorie de la Sp\'eculation, {\em Annales de l'Ecole Normale Sup\'erieure}  (1900).

\bibitem{bs1} F. Black and M. Scholes, The Pricing of Options and Corporate Liabilities,  {\em Journal of Political Economy} {\bf 81} (1973) 637-45.

\bibitem{bs2} R. C Merton, Theory of Rational Option Pricing,
{\em Bell Journal of Economics} {\bf 4} (1973) 141-183.

\bibitem{cir} J.C. Cox, J.E. Ingersoll, and S.A. Ross,
A Theory of the Term Structure of Interest Rates,  
{\it Econometrica} {\bf 53} (1985) 385-408.

\bibitem{dupire} B. Dupire, Pricing with a Smile, {\em Risk} {\bf 7} 18-20 (1994).

\bibitem{vasicek} O. Vasicek,
An equilibrium Characterization of the Term Structure,
{\it Journal of Financial Economics} {\bf 5} (1977) 177-188.

\bibitem{aitsahalia} Y. A\"{\i}t Sahalia, 
Transition Densities for Interest Rate and Other Nonlinear Diffusions,
{\it Journal of Finance} {\bf 54} (1999) 1361-1395.

\bibitem{makri} N. Makri and W.H. Miller,
Exponential Power Series Expansion for the Quantum Time Evolution Operator,
{\it Journal of Chemical Physics} {\bf 90} (1989) 904-911.

\bibitem{rosaclot1} M. Bennati, M. Rosa-Clot, and S. Taddei,
A Path Integral Approach to Derivative Security Pricing I: 
General Method and Analytical Results,
{\it International Journal of Theoretical and Applied Finance} {\bf 2}
(1999) 381-407. 

\bibitem{rosaclot2} M. Rosa-Clot, and S. Taddei,  
A Path Integral Approach to Derivative Security Pricing II: 
Numerical Results,
{\it International Journal of Theoretical and Applied Finance} {\bf 5}
(2002) 123-146. 

\bibitem{shreve} S. E. Shreve, {\em Stochastic Calculus for Finance} (Springer-Verlag, New York, 2004).

\bibitem{pireference1} R. P. Feynman,
Space-Time Approach to Non-Relativistic Quantum Mechanics,
{\em Review of Modern Physics} {\bf 20} (1948) 367-387. 

\bibitem{pireference2} R. P. Feynman and A. R. Hibbs, 
{\em Quantum Mechanics} (McGraw-Hill, New York, 1965).

\bibitem{montagna1} G. Montagna, O. Nicrosini, and N. Moreni, 
A Path Integral Way to Option Pricing,
{\it Physica A} {\bf 310} (2002) 450-466. 

\bibitem{montagna2} G. Bormetti, G. Montagna, N. Moreni, and N. Nicrosini, 
Pricing Exotic Options in a Path Integral Approach,
{\em Quantitative Finance} {\bf 6} (2006) 55-66.

\bibitem{baaquiep} B.E. Baaquie, C. Corian\`o, and M. Srikant,  
Hamiltonian and Potentials in Derivative Pricing Models: 
Exact Results and Lattice Simulations,
{\it Physica A} {\bf 334} (2004) 531-557.

\bibitem{matacz} A. Matacz,
Path Dependent Option Pricing: the Path Integral Partial Averaging Method,
e.print arXiv:cond-mat/0005319 (2000).  

\bibitem{dash} J. W. Dash, {\em Quantitative Finance and Risk Management: a Physicist's approach} (World Scientific, Singapore, 2004).


\bibitem{cev} J. C. Cox and S. A. Ross, 
The Valuation of Options for Alternative Stochastic Processes,
{\em Journal of Financial Economics} {\bf 3} (1976) 145-166.

\bibitem{feller} W. Feller,
The Parabolic Differential Equations and the Associated Semi-Groups of Transformations, 
{\it Annals of Mathematics} {\bf 55} (1952) 468-519. 

\bibitem{abramovitz} M. Abramovitz, I. A. Stegun, {\em Handbook of Mathematical Functions} (National Bureau of Standards, Applied Mathematics Series, 1965).

\bibitem{makri2} N. Makri and W.H. Miller, 
Correct Short Time Propagator for Feynman Path Integration by Power Series 
Expansion in $\Delta t$, 
{\it Chemical Physics Letters} {\bf 151} (1989) 1-8.

\bibitem{baaquie} B. E. Baaquie, {\em Quantum Finance} (Cambridge University Press, 2004).

\bibitem{lin} V. Linetsky,
The Path Integral Approach to Financial Modeling and Options Pricing,
{\it Computational Economics} {\bf 11} (1998) 129-163.

\bibitem{schulman} L.S. Schulman, {\em Techniques and applications of path integration} (Wiley, New York, 1981).

\bibitem{quasifourier} R.D. Coalson, 
On the Connection between Fourier Coefficient and Discretized Cartesian 
Path Integration,
{\it Journal of Chemical Physics} {\bf 85} (1986) 926-936. 

\bibitem{metropolis} N. Metropolis, A.W. Rosenbluth, M.N. Rosenbluth, H. Teller, and E. Teller, 
Equation of State Calculations by Fast Computing Machines,
{\it Journal of Chemical Physics} {\bf 21} (1953) 1087-1092. 

\bibitem{glassermann} P. Glasserman, {\em Monte Carlo Methods in Financial Engineering} (Springer-Verlag, New York, 2004).

\bibitem{lucaump} L. Capriotti, {\em in preparation}.

\end{thebibliography}
\end{document}